\documentstyle[prd,preprint,aps,floats,epsfig]{revtex}
\draft
\tighten
\newcommand{\beq}{\begin{equation}}
\newcommand{\eeq}{\end{equation}}
\newcommand{\bea}{\begin{eqnarray}}
\newcommand{\eea}{\end{eqnarray}}

\begin{document}

\title{Single spin asymmetry in DVCS}

\author{Andreas Freund$^a$, Mark Strikman$^b$}

\address{$^a$ I.N.F.N, Sezione Firenze, Lg. Enrico Fermi 2, 50125 Florence, Italy\\
$^b$ The Pennsylvania State University, University Park, Pa 16802, USA and DESY-Hamburg, Germany}

\maketitle

\begin{abstract}
In the following note, we will present an estimation of the single spin 
asymmetry in deeply virtual Compton scattering (DVCS) which directly
allows one to test predictions of the ratio of the imaginary part of the
amplitude in DIS to DVCS, as well as access the skewed parton distributions
at small $x$ in the DGLAP region. We find it to be large for the HERA kinematics 
to be accessible in forthcoming
runs with polarized electrons. \newline

\end{abstract}

\section{Introduction}
\label{intro}

The single spin asymmetry (SSA) in deeply virtual Compton scattering (DVCS)
which was first discussed in Ref.\ \cite{1} and subsequently in \cite{2,2a}, 
is 
directly proportional to the interference term between the DVCS and the 
Bethe-Heitler (BH) process and thus offers
a direct way of accessing the imaginary part of the DVCS amplitude 
\cite{1,2,2a}
 in the scattering
 of polarized electrons off unpolarized protons.
Such measurements would be feasible at HERA in the near future
(around the end of 2000) when the spin rotators will be installed both at 
the ZEUS and H1 experiment.
This then allows a direct check of our prediction
\cite{3} obtained in the $\alpha_s~lnQ^2$ approximation of a large (a factor
$\sim 2$)
enhancement  of the imaginary part of the DVCS amplitude for
small $x$ kinematics as compared to the
 the imaginary part of the DIS amplitude as measured through 
$F_2(x,Q^2)$. Furthermore, this asymmetry is 
another physical observable, besides the azimuthal angle asymmetry $A$ proposed in
\cite{3},  which allows one to directly access and extract 
the skewed parton distributions (SPD's) \cite{4}, at least in the DGLAP region \cite{5}.
This note  is structured in the following way: In the next section we will 
give the
definition of the SSA along with an analytical expression to $\alpha_s~\ln Q^2$
accuracy and in Sec.\ \ref{numb} we will give numbers for this asymmetry in
the HERA kinematical regime, followed by conclusions.

\section{The Single Spin Asymmetry}
\label{ssa1}

In order to form the SSA in DVCS one scatters polarized electrons/positrons off
an unpolarized proton target and then takes the difference in differential 
cross sections between positive and 
negative helicity initial states. The exact definition is
\beq
A_s = \frac{\int_{0}^{\pi}d\phi~(d\sigma_{+}-d\sigma_{-})_{DVCS+BH} - 
\int_{\pi}^{2\pi}d\phi~(d\sigma_{+}-d\sigma_{-})_{DVCS+BH}}
{\int_{0}^{2\pi}d\phi~(d\sigma_{+}+d\sigma_{-})_{DVCS+BH}}  
\label{defasym}
\eeq
where $+,-$ refers to the helicity state, $\phi$ is the azimuthal angle in the 
transverse scattering plane as defined in \cite{3} and $DVCS+BH$ means that we 
are considering the total differential cross section of these two processes.
Note that $d\sigma_{DVCS+BH,unpol.} = \frac{1}{2} (d\sigma_{+} + d\sigma_{-})$. 
There is another way of writing the SSA as well as $A$ 
by weighting the respective interference term by $sin(\phi)$ in
the case of SSA and $cos(\phi)$ in the case of $A$, as originally proposed by 
the authors of \cite{2a}, so one obtains in the SSA case 
\beq
A_s = \frac{\int_{0}^{2\pi}d\phi~\sin (\phi)(d\sigma_{+}-d\sigma_{-})_{DVCS+BH}}
{\int_{0}^{2\pi}d\phi~(d\sigma_{+}+d\sigma_{-})_{DVCS+BH}}.  
\label{defasym1}
\eeq
The difference between this and our definition is a factor of $\frac{\pi}{4}$ by
which the asymmetry from Eq.\ (\ref{defasym1}) would be smaller than the one
from Eq.\ (\ref{defasym}). The advantage of the second definition might be that 
one does not need to know whether the final state electron
and photon are in the same or opposite detector hemispheres, one just has to 
integrate all the data by the respective weight function.

In the center of mass frame of the final state photon and proton, we find the
following expression for $d\sigma_{int} = d\sigma_{+}-d\sigma_{-}$ following
our methods from \cite{3} for small $x,t$ and $Q^2>>-t$ and the results
for the unpolarized hadronic and spin-dependent leptonic tensor for the
interference part in the form given in \cite{2}:
\bea
\frac{d\sigma_{int}}{dxdyd|t|d\phi} &=& \frac{2\alpha^3s_{ep}F_2(x,Q^2)F_1(|t|)e^{-B|t|/2}sin(\phi)y}
{\sqrt{|t|}Q^5R\sqrt{1-y}}\nonumber\\
& &\left ( 1 - (1-y)^2 + \frac{2\sqrt{|t|}cos(\phi)}{Q}\frac{1-(1-y)^3}{\sqrt{1-y}} \right )
\label{inteq}
\eea
where $R$ is the ratio of the imaginary part of the DIS to the imaginary part of 
the DVCS amplitude which was taken from \cite{3}, $x,y,t$ and $Q^2$ are the usual 
kinematical
invariants, $\phi$ is the azimuthal angle of the leptonic and hadronic scattering plane \cite{3}
specifying the off-planarity of the event,
$s_{ep}$ is the center of mass energy of the electron-proton 
system, $F_2(x,Q^2)$ the normal structure function \cite{3a}, $B$ is the 
slope of the $t$-dependence of the DVCS amplitude which we took to be an 
exponential at small $t$ and
\beq
F_1(|t|) = \frac{G_E(|t|) + \frac{|t|}{4M^2}G_M(|t|)}{1+\frac{|t|}{4M^2}}
\eeq
the t-dependence of the BH contribution appearing in the interference term
\cite{3}, where we use the regular dipole fit for the electric and magnetic 
nucleon form factors. Note that the term $sin(\phi)cos(\phi)$ in Eq.\ 
(\ref{inteq}) integrates to zero in the definition of $A_s$. Eq.\ (\ref{inteq})
is in agreement with Ref.\ \cite{2}, where however the
DVCS amplitude was not computed. 
For the total, unpolarized, differential cross section in 
the denominator of $A_s$ we use our results from \cite{3}
\bea 
\frac{d\sigma_{DVCS+BH,\mbox{unpol.}}}{dxdyd|t|d\phi}&=&\frac{\pi\alpha^3s}{4R^2Q^6}(1+(1-y)^2)
e^{-B|t|}F^2_2(x,Q^2)(1+\eta^2)\nonumber\\
&+& \frac{\alpha^3 s y^2(1+(1-y)^2)}{\pi Q^4
|t| (1-y)}\left [ \frac{G_E^2(t) + \tau G_M^2(t)}{1+\tau} \right ]\nonumber\\
&+& \frac{\eta \alpha^3 s y(1+(1-y)^2) cos(\phi) e^{-B|t|/2} F_2(x,Q^2)}{2 Q^5 \sqrt(|t|)
\sqrt(1-y) R}\left [ \frac{G_E(t) + \tau G_M(t)}{1+\tau} \right ]
\eea
where the first term corresponds to the DVCS differential cross section,
the second to the BH differential cross section and the third term is the interference 
contribution of DVCS and BH to the total differential cross section and $\eta$ is the
ratio of real to imaginary part of the DIS amplitude \cite{3}. Note that after 
the $\phi$ integration as required in the denominator of Eq.\ (\ref{defasym}), the
interference term drops out and one is only left with the DVCS and BH contributions 
respectively.

\section{Numbers for the Single Spin Asymmetry}
\label{numb}

At $Q^2=3.5~\mbox{GeV}^2$ with a $B$ of $8~\mbox{GeV}^{-2}$, we find the 
asymmetry $A_s$ from Eq.\ (\ref{defasym}) to be maximal around $32\%$ at  
$y=0.5$ and  
$-t=0.2~\mbox{GeV}^2$ for $x=10^{-4}$ and about $27\%$ for $x=10^{-3}$.

\begin{figure}
\vskip-1in
\centering
\epsfig{file=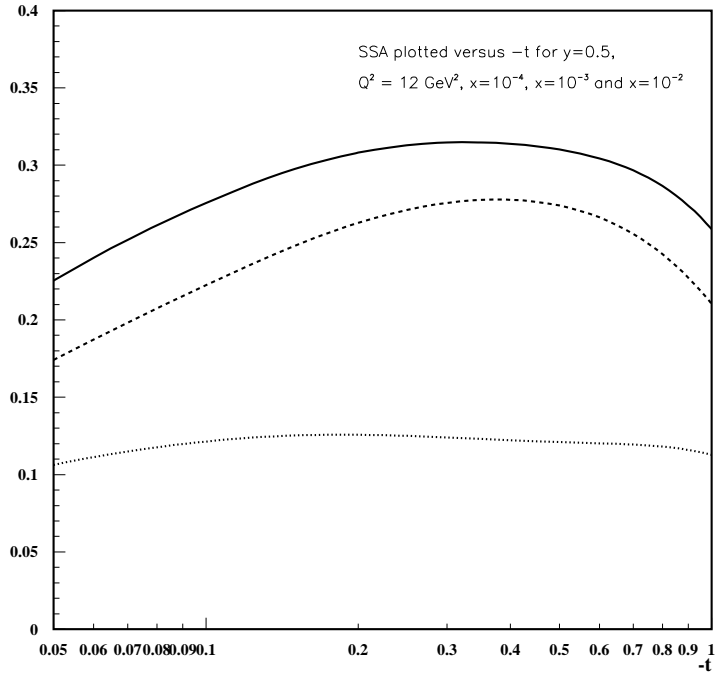,height=12cm}
\vskip-0.9in
\epsfig{file=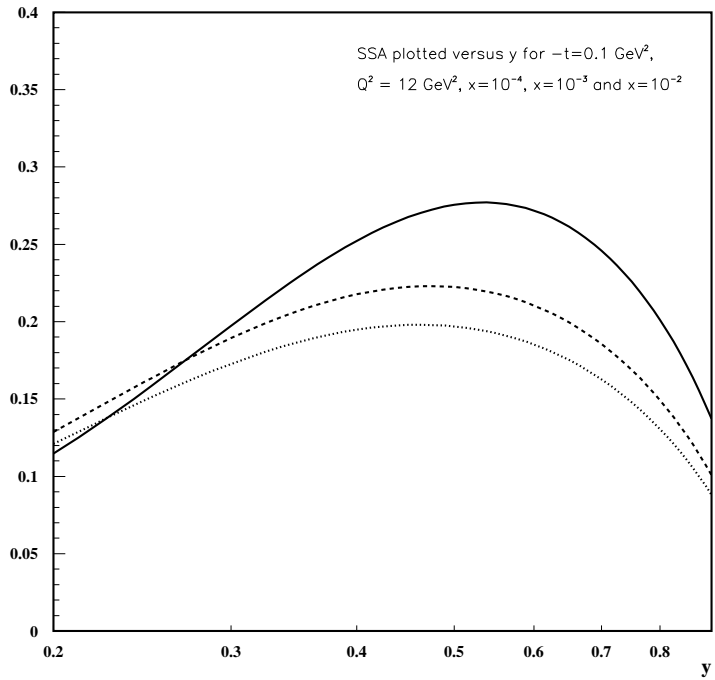,height=12cm}
\vspace*{5mm}
\vskip-0.5in
\caption{a) The SSA is plotted versus $-t$ for 
$x=10^{-4}$ (solid curve), 
$x=10^{-2}$ (dotted curve) and $x=10^{-3}$ (dashed curve)
for $Q^2=12~\mbox{GeV}^2$, $B=5~\mbox{GeV}^{-2}$ and $y=0.5$.
b) The SSA is plotted versus $y$ for the same $x,Q^2,B$ and 
$-t=0.1~\mbox{GeV}^{2}$.}
\label{tdep4}
\end{figure}

\begin{figure}
\vskip-1in
\centering
\epsfig{file=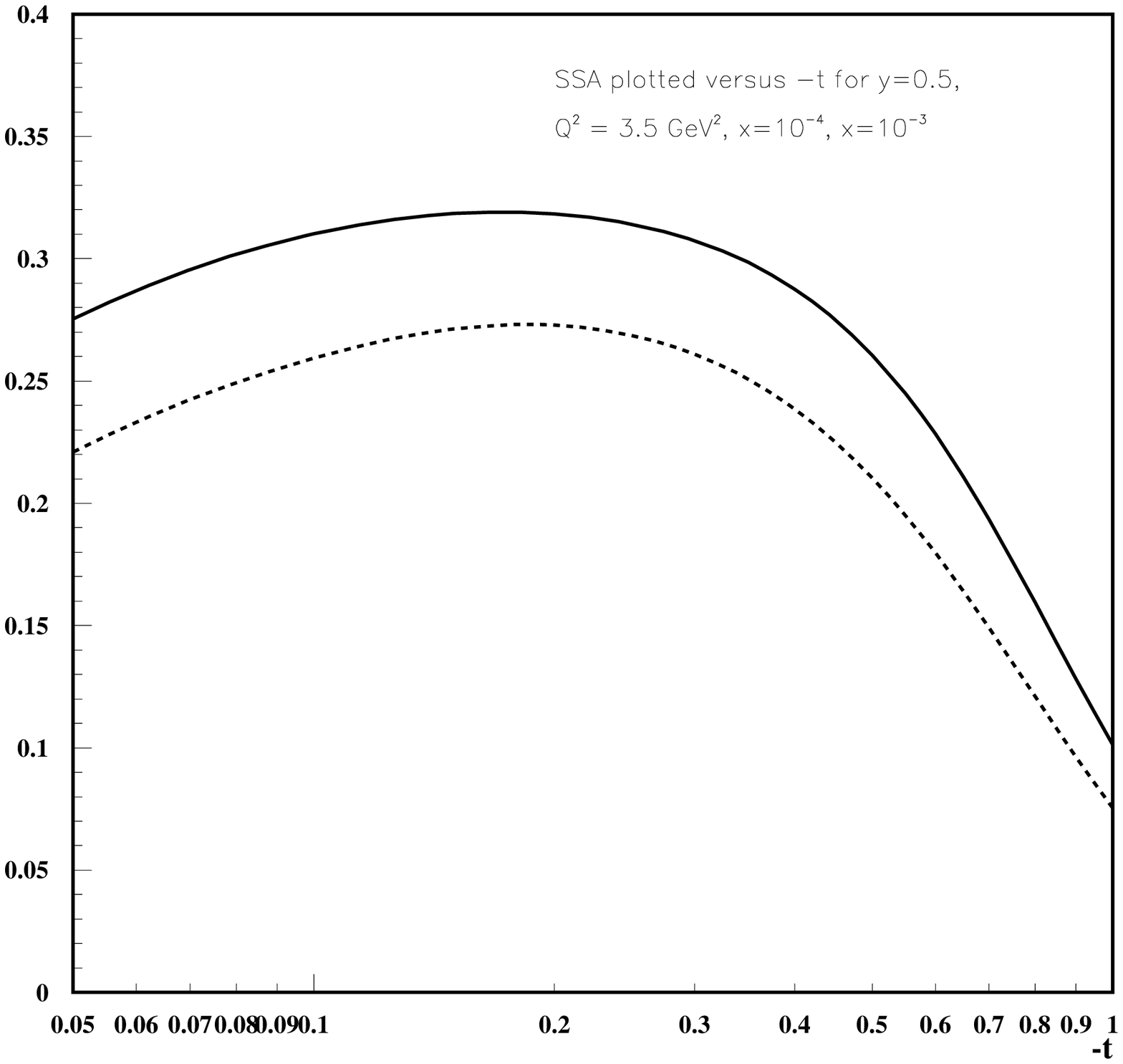,height=12cm}
\vskip-0.9in
\epsfig{file=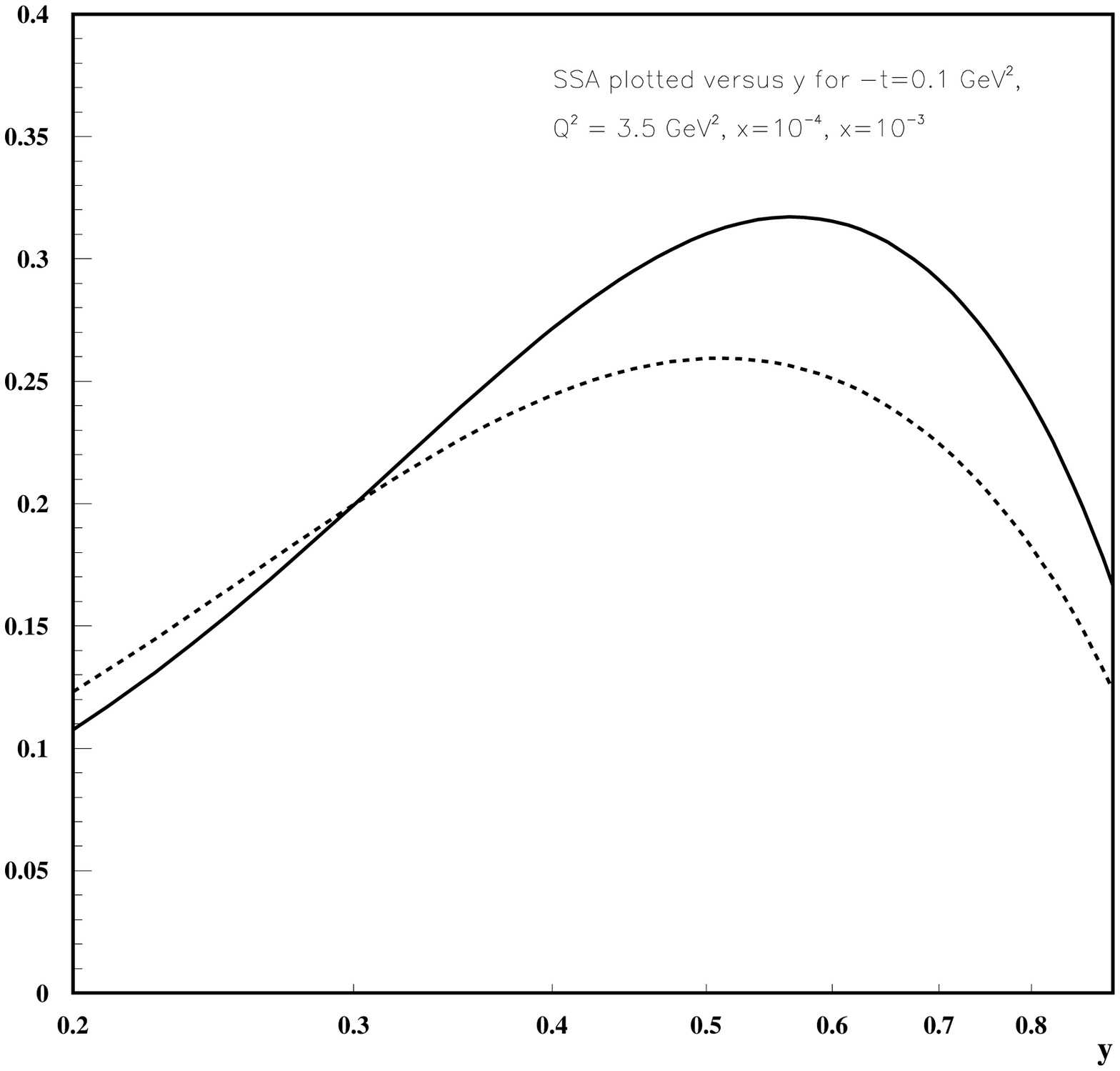,height=12cm}
\vspace*{5mm}
\vskip-0.5in
\caption{a) The SSA is plotted versus $-t$ for 
$x=10^{-4}$ (solid curve),  and $x=10^{-3}$ (dashed curve)
for $Q^2=3.5~\mbox{GeV}^2$, $B=8~\mbox{GeV}^{-2}$ and $y=0.5$.
b) The SSA is plotted versus $y$ for the same $x,Q^2,B$ and 
$-t=0.1~\mbox{GeV}^{2}$.}
\label{tdep5}
\end{figure}

At  $Q^2=12~\mbox{GeV}^2$ with a $B$ of $5~\mbox{GeV}^{-2}$, we find the 
asymmetry $A_s$ from Eq.\ (\ref{defasym}) to be maximal around $31\%$ at  
$y=0.5$ and  $-t=0.35~\mbox{GeV}^2$ for $x=10^{-4}$ and about $28\%$ for 
$x=10^{-3}$ and about $12\%$ at $x=10^{-2}$. For more details see 
Figs.\ \ref{tdep4} and \ref{tdep5}.


This very large asymmetry, in comparison with the azimuthal angle asymmetry $A$
from \cite{3}, is mainly due to the fact that first, we are dealing with the
imaginary part of the DVCS amplitude which is about a factor $3$ to $4$ larger than 
the real part and secondly one should note that, despite the almost identical 
structure of terms in $A_s$ and $A$, there is an extra factor
of $2$ in $A_s$ as compared to $A$. To counter these huge enhancement factors 
there is only the altered $y$ dependence, albeit quite strongly altered, which
pushes the maximum of $A_s$ to larger $y$ values than in the case of $A$ and 
forces $A_s$ only to be about a factor $2$ to $3$ larger than $A$.

Experimentally, such a huge asymmetry should be fairly easily measurable at HERA, 
which
as mentioned in Sec.\ \ref{intro}, would give one an interesting opportunity 
to compare the imaginary parts of the DIS and DVCS amplitudes as well as being
able to extract the skewed gluon distribution at small $x$ albeit only
in the DGLAP region.    

\label{Conclusions}

In this note we have shown that the single spin asymmetry in DVCS should be very 
sizable in the HERA regime of moderately large $Q^2$, small $x$ and small $t$ 
and thus fairly easily measurable at HERA. This opens new avenues of comparing
hadronic amplitudes as well as extracting the skewed gluon distribution at 
small $x$ in the DGLAP region. 

\section*{Acknowledgements}

A.F was supported by the E.\ U.\ contract $\#$FMRX-CT98-0194 and M.S was 
supported in part by the U.S. Department of
Energy.


\begin{references}
\bibitem{5} Note that the imaginary part of the DVCS amplitude is only 
            sensitive to the SPD's in the DGLAP region, see for example
            \cite{2,2a,3,4}.
\end{references}
\end{document}